\documentclass{wscpaperproc}
\usepackage{latexsym}
\usepackage{graphicx}
\usepackage{mathptmx}
\usepackage[T1]{fontenc}

\usepackage{amsmath}
\usepackage{amsfonts}
\usepackage{amssymb}
\usepackage{amsbsy}
\usepackage{amsthm}
\usepackage{float, caption}
\usepackage[pdftex,colorlinks=true,urlcolor=blue,citecolor=black,anchorcolor=black,linkcolor=black]{hyperref}

\newtheoremstyle{wsc}% hnamei
{3pt}% hSpace abovei
{3pt}% hSpace belowi
{}% hBody fonti
{}% hIndent amounti1
{\bf}% hTheorem head fontbf
{}% hPunctuation after theorem headi
{.5em}% hSpace after theorem headi2
{}% hTheorem head spec (can be left empty, meaning `normal')i

\theoremstyle{wsc}

\begin{document}

%***************************************************************************
% AUTHOR: AUTHOR NAMES GO HERE
% FORMAT AUTHORS NAMES Like: Author1, Author2 and Author3 (last names)
%
%		You need to change the author listing below!
%               Please list ALL authors using last name only, separate by a comma except
%               for the last author, separate with ''and''
%
% \WSCpagesetup{LastName1, LastName2, LastName3, LastName4, and LastName (LastAuthor)}

% setting up general page style
\pagestyle{fancyplain}

% setting up page style of first page
\thispagestyle{plain}
\firstPageHead{}

% setting up running header (authors) of subsequent pages
\chead{\fancyplain{}{\itshape Camur, Tseng, Thanos,  White, Yund, and Iakovou}}

% setting up seperation parameters
%\headsep=72pt
\rhead{}
\cfoot{}
\renewcommand{\headrulewidth}{0pt} % (renewcommand needed in fancyhdr to remove top decorative line)
%\headrulewidth=0pt  % ("setlength" needed in fancyheading to remove top decorative line)

%%%%%%%%%%%%%%%%%%%%%%%%%%%%%%%%%%%%%%%%%%%%%%%%%%%%%%%%%%%%%%%%%%%%%%%%%%%%%%
%                                                                            %
%     THESE COMMANDS ARE REQUIRED TO WORK WITH WSC.BST TO MAKE BIBLIO     %
%                                                                            %
%%%%%%%%%%%%%%%%%%%%%%%%%%%%%%%%%%%%%%%%%%%%%%%%%%%%%%%%%%%%%%%%%%%%%%%%%%%%%%
\makeatletter
\let\@internalcite\cite
\def\cite{\def\@citeseppen{-1000}%
    \def\@cite##1##2{(##1\if@tempswa , ##2\fi)}%
    \def\citeauthoryear##1##2##3{##1 ##3}\@internalcite}
\def\citeNP{\def\@citeseppen{-1000}%
    \def\@cite##1##2{##1\if@tempswa , ##2\fi}%
    \def\citeauthoryear##1##2##3{##1 ##3}\@internalcite}
\def\citeN{\def\@citeseppen{-1000}%
%  Pierre L'Ecuyer's fix for multiple cite bug
%  Added by Paul J Sanchez on 4 October 2001
%   \def\@cite##1##2{##1\if@tempswa , ##2)\else{)}\fi}%
%   \def\citeauthoryear##1##2##3{##1 (##3}\@citedata}
    \def\@cite##1##2{##1\if@tempswa, ##2)\else{}\fi}%
    \def\citeauthoryear##1##2##3{##1 (##3)}\@citedata}
\def\citeA{\def\@citeseppen{-1000}%
    \def\@cite##1##2{(##1\if@tempswa , ##2\fi)}%
    \def\citeauthoryear##1##2##3{##1}\@internalcite}
\def\citeANP{\def\@citeseppen{-1000}%
    \def\@cite##1##2{##1\if@tempswa , ##2\fi}%
    \def\citeauthoryear##1##2##3{##1}\@internalcite}
\def\shortcite{\def\@citeseppen{-1000}%
    \def\@cite##1##2{(##1\if@tempswa , ##2\fi)}%
    \def\citeauthoryear##1##2##3{##2 ##3}\@internalcite}
\def\shortciteNP{\def\@citeseppen{-1000}%
    \def\@cite##1##2{##1\if@tempswa , ##2\fi}%
    \def\citeauthoryear##1##2##3{##2 ##3}\@internalcite}
\def\shortciteN{\def\@citeseppen{-1000}%
%  Pierre L'Ecuyer's fix for multiple cite bug
%  Added by Paul J Sanchez on 2 September 2002
%  should have caught this last year...
%   \def\@cite##1##2{##1\if@tempswa , ##2)\else{)}\fi}%
%   \def\citeauthoryear##1##2##3{##2 (##3}\@citedata}
% Shane G. Henderson fix for extra right bracket at end of optional material June 8, 2005
%    \def\@cite##1##2{##1\if@tempswa, ##2)\else{}\fi}%
    \def\@cite##1##2{##1\if@tempswa, ##2\else{}\fi}%
    \def\citeauthoryear##1##2##3{##2 (##3)}\@citedata}
\def\shortciteA{\def\@citeseppen{-1000}%
    \def\@cite##1##2{(##1\if@tempswa , ##2\fi)}%
    \def\citeauthoryear##1##2##3{##2}\@internalcite}
\def\shortciteANP{\def\@citeseppen{-1000}%
    \def\@cite##1##2{##1\if@tempswa , ##2\fi}%
    \def\citeauthoryear##1##2##3{##2}\@internalcite}
\def\citeyear{\def\@citeseppen{-1000}%
    \def\@cite##1##2{(##1\if@tempswa , ##2\fi)}%
    \def\citeauthoryear##1##2##3{##3}\@citedata}
\def\citeyearNP{\def\@citeseppen{-1000}%
    \def\@cite##1##2{##1\if@tempswa , ##2\fi}%
    \def\citeauthoryear##1##2##3{##3}\@citedata}
%
% \@citedata and \@citedatax:
%
% Place commas in-between citations in the same \citeyear, \citeyearNP,
% \citeN, or \shortciteN command.
% Use something like \citeN{ref1,ref2,ref3} and \citeN{ref4} for a list.
%
\def\@citedata{%
    \@ifnextchar [{\@tempswatrue\@citedatax}%
                  {\@tempswafalse\@citedatax[]}%
}

\def\@citedatax[#1]#2{%
\if@filesw\immediate\write\@auxout{\string\citation{#2}}\fi%
  \def\@citea{}\@cite{\@for\@citeb:=#2\do%
    {\@citea\def\@citea{, }\@ifundefined% by Young
       {b@\@citeb}{{\bf ?}%
       \@warning{Citation `\@citeb' on page \thepage \space undefined}}%
{\csname b@\@citeb\endcsname}}}{#1}}%

% don't box citations, separate with ; and a space
% also, make the penalty between citations negative: a good place to break.
%
\def\@citex[#1]#2{%
\if@filesw\immediate\write\@auxout{\string\citation{#2}}\fi%
  \def\@citea{}\@cite{\@for\@citeb:=#2\do%
    {\@citea\def\@citea{; }\@ifundefined% by Young
       {b@\@citeb}{{\bf ?}%
       \@warning{Citation `\@citeb' on page \thepage \space undefined}}%
{\csname b@\@citeb\endcsname}}}{#1}}%

% (from apalike.sty)
% No labels in the bibliography.
%
\def\@biblabel#1{}
\makeatother

%\newlength{\bibhang}
%\setlength{\bibhang}{2em}

% Indent second and subsequent lines of bibliographic entries. Taken
% from openbib.sty: \newblock is set to {}.
% \renewcommand{\refname}{REFERENCES}

\newdimen\bibindent
\bibindent=0.0em
% SEC: was \def\thebibliography#1{\section*{\refname\@mkboth
% SEC: was   {\uppercase{\refname}}{\uppercase{\refname}}}\list
\def\thebibliography#1{\section*{\refname}\list
   {}{\settowidth\labelwidth{[#1]}
   \leftmargin\parindent
   \itemindent -\parindent
   \listparindent \itemindent
   \itemsep 0pt
   \parsep 0pt}
   \def\newblock{}
   \sloppy
   \sfcode`\.=1000\relax}

           % Set up BiBTeX macros

% needed to make the tex document look more like the word counterpart :-(
\setlength{\baselineskip}{12.7pt}

% AUTHOR: Enter the title, all letters in upper case
\title{AN INTEGRATED SYSTEM DYNAMICS AND DISCRETE EVENT SUPPLY CHAIN SIMULATION FRAMEWORK FOR SUPPLY CHAIN RESILIENCE WITH NON-STATIONARY PANDEMIC DEMAND}

% AUTHOR: Enter the authors of the article, see end of the example document for further examples
% \author{Canan Gunes Corlu\\[12pt]
% 	Metropolitan College\\
% 	Boston University\\
% 	1010 Commonwealth Avenue\\
% 	Boston, MA 02215, USA\\
% % Multiple authors are entered as follows.
% % You may also need to adjust the titlevbox size in the preamble - search for titlevboxsize
% \and
% Susan R. Hunter\\[12pt]
% School of Industrial Engineering\\
% Purdue University\\
% 315 N Grant Street\\
% West Lafayette, IN 47906, USA\\
% \and
% Henry Lam\\ [12pt]
% Industrial Engineering and\\
% Operations Research\\
% Columbia University\\
% 500 West 120th Street\\
% New York, NY 10027, USA\\
% \and
% Bhakti Stephan Onggo\\ [12pt]
% CORMSIS - Southampton Business School\\
% University of Southampton\\
% University Road\\
% Southampton, SO17 1BJ, UK\\
% }

% \WSCpagesetup{Camur, Tseng, Thanos,  White, Yund, Iakovou}

% AUTHOR: Enter the authors of the article, see end of the example document for further examples
\author{Mustafa Can Camur\\
$[$Aristotelis E. Thanos$]$ \\ 
$[$Walter Yund$]$\vspace{12pt} \\
Optimization $\&$ Risk Analysis Team \\
GE   Research Center\\
Niskayuna, NY 12309 USA\\
 \and 
Chin-Yuan  Tseng\\ 
$[$Chelsea C. White$]$ \vspace{12pt}\\ 
Department of Industrial and Systems Engineering \\
Georgia Institute of Technology\\
Atlanta, GA 30332 USA\\
\and
Eleftherios  Iakovou\\ [12pt]
Department of Mechanical Engineering \\
 Texas A$\&$M University\\
College Station, TX 77843 USA\\
}
\maketitle
\section*{ABSTRACT}
COVID-19 resulted in some of the largest supply chain disruptions in recent history. To mitigate the impact of future disruptions,  an integrated hybrid simulation framework is proposed to couple nonstationary demand signals from an event like COVID-19 with a model of an end-to-end supply chain. First, a system dynamics susceptible-infected-recovered (SIR) model is created, augmenting a classic epidemiological model to create a realistic portrayal of demand patterns for oxygen concentrators (OC). Informed by this granular demand signal, a supply chain discrete event simulation model of OC sourcing, manufacturing, and distribution is  developed to test production augmentation policies to satisfy this increased demand. This model utilizes publicly available data, engineering teardowns of OCs, and a supply chain illumination to identify suppliers. The findings indicate that this coupled approach can use realistic demand during a disruptive event to enable rapid recommendations of policies for increased supply chain resilience with controlled cost.

\section{INTRODUCTION}
\label{sec:intro}
The recent reaction to the COVID-19 pandemic  demonstrated the fragility of supply chains to disruptions of supply, and corresponding difficulty in reacting to non-stationary demand (i.e., exponential growth in demand over a short period of time). In a survey, the majority of manufacturing respondents reported increased lead times and lead time churn as a result of the COVID-19 pandemic \cite{boyd2020COVID}. %Global economic problems including high inflation rates started arising due to the supply-demand disparity \shortcite{de2023demand}.
In addition, freight networks were disrupted causing delays in supply chain operations \shortcite{camur2022optimization}.  Coupling these supply disruptions to equally significant demand shocks, the result has been significant material shortages in wide-ranging product categories, from healthcare supplies and equipment to basic goods. 

In response to these linked shocks, the Robotics and Automation Decision Framework for Agility and Resilience (RADAR) project is developing a decision system for deploying robotics and automation in a stressed supply chain in order to mitigate the effects of these shocks, and enhance supply chain agility, resilience, and preparedness. The RADAR framework addresses supply chain competitiveness and resilience through three core thrusts: macro-scale modeling of a pandemic-related supply chain, micro-scale modeling of robotics within a manufacturing facility, and physical demonstration of enhanced response. This paper plays a significant role in the  project by contributing to the macro- and micro-scale modeling components of the proposed end-to-end framework.

To demonstrate this framework, the RADAR project selected Oxygen Concentrators (OCs) as a manufacturing and supply chain use case, due to both the COVID demand shocks that affected OC supply chains, as well as the opportunity to augment OC manufacturing with the rapid introduction of robotics in manufacturing facilities.  OCs are medical devices that filter nitrogen from the air and  provide a higher amount of oxygen to a patient. Both hospitalized and home patients may require oxygen support due to potential respiratory complications after being infected by COVID-19. The demand for OCs surged exponentially during the pandemic, while also experiencing supply disruptions. OCs were reported as scarce health resources in many states in the U.S. \shortcite{devereaux2023oxygen}.  Thus, it is vital to understand the demand behavior for OCs and how / where those demands may be met during a healthcare crisis. Similar models can be developed for other critical healthcare supplies, provided data on market size and pandemic usage are available.

While significant disruptions such as the COVID-19 pandemic are inevitable, an over-reliance on standard supply chain models of consumption and lead time, which do not take into account the probability and impact of these significant disruptions, leave these models exposed to significant shortcomings during a non-stationary event.  Additionally, models which may assist in helping supply chain managers react to these events (e.g., epidemiological models), tend to suffer from an over-reliance on expert assumptions.  The result is a fragmented approach that fails to ensure sufficient supplies are available during a disruptive event. Thus, there has been significant interest in three fundamental simulation-based modeling approaches to understand the spread of COVID-19 in communities as discussed below. 
\begin{enumerate}
    \item  Susceptible-Infected-Removed (SIR) models: This is the most common methodology that groups a population of a specific region into different categories (i.e. Susceptible, Exposed, Infected, and Recovered) and applies mathematical ratios or rules about how individuals in the population move from one category to another using scientific assumptions about the disease \shortcite{salimipour2023sir}. 
    %\shortcite{cooper2020sir,salimipour2023sir}. 
    \item  Extrapolation models: This methodology infers trends about a pandemic in a specific area by observing the historical and current state of the spread and then applying an estimate of the possible future pandemic spread path, while also using information from other locations with similar characteristics \shortcite{ho2023go}. 
    %\shortcite{jiang2023time,ho2023go}. 
    \item  Agent-based models: This modeling approach is based on the bottoms up creation of a simulated population and follow interactions among individuals, called agents, in that region, based on characteristics, rules, behavior, movement, mixing patterns, risks, intervention policies, and social networks (work, family, transportation, social interaction, buying patterns etc.) \shortcite{shamil2021agent}. %\shortcite{truszkowska2021high,shamil2021agent}. 
\end{enumerate}

These models either i) assume that key features of a target disease are known and stationary with high certainty (e.g., infectivity, mortality rate), or ii) use historical data to project/extrapolate future trajectories. During the COVID-19 pandemic, the trajectory of cases has been unique in different regions even within the same country \shortcite{sapkota2021chaotic}, making a dynamic SIR model with elements of agent-based modeling, particularly in hospitalizations, the best choice for modeling the COVID-19 pandemic. In this context, creating a unified approach to tying models of a disruptive event (e.g., pandemic, natural disaster, humanitarian crisis, war) with that event's impact on a supply chain, will enable decision-makers to minimize the risk of material shortage, ensuring a fully functioning supply chain network. Further, robust design and carefully crafted output of these simulation models enable future integration into optimization models to make sophisticated and accurate policy decisions \shortcite{vogiatzis2019identification,camur2022stochastic}.

 Current epidemiological models (and other anticipatory intelligence models aimed at predicting grey swan events) are generally aimed at informing policy decisions  and have little influence on supply chain response until such events are already underway.  This results in a supply chain response that is reactive instead of proactive.  Additionally, current inventory management, procurement, and supply chain practices within hospital networks assume stationary, possibly seasonal demand, and do not typically use epidemiological models \shortcite{ali2021supply}.

 A hybrid simulation framework that integrates two simulation models is proposed to understand i) the realistic demand for OCs that are needed by COVID-19 patients, and ii) how supply chain operations shall be conducted to meet this demand surge. The contribution includes the design and implementation of a data-driven decision support framework. This framework captures the realistic baseline of a supply chain network and the effects of a major disruption event, such as a pandemic, and allows the experimentation of alternative scenarios in several aspects of the model (demand signal, supply chain disruption, change of policies, alternate suppliers, etc.) without modification of source code. The technical novelty can be summarized as follows: i) creation of a high-fidelity supply chain model containing advanced demand predictions that come from a pandemic model tuned to represent accurately historical captured data, ii) utilization of multiple simulation methodologies including system dynamics, agent-based modelling and discrete event simulation to incorporate the advantages of each of these techniques, iii) adoption of a data-driven approach that enables handling multiple alternatives with the same model without modifying the source code and  iv) creation of a multi-scale supply chain simulation model utilizing supplier illumination.

\section{SIR EPIDEMIOLOGY MODEL} \label{sec:SIR}

In the classical SIR models, it is assumed that the parameters of the epidemiological model are kept constant, i.e. total population remains the same throughout stages of a pandemic, the susceptible population converges to zero and the reproduction ratio ($R_0$) is constant or near-constant  \shortcite{cooper2020sir}. However, COVID-19 pandemic has followed different characteristics: mitigation policies and multiple reinfections make the classical system dynamics approach insufficient, mostly because of the large fluctuations of $R_0$ \shortcite{moein2021inefficiency}. To move beyond these limitations, a modified SIR model that aims to incorporate the unique spread of the virus is proposed and  realistic demand for OCs in each state is modelled. To this end the model incorporates dynamically changing epidemiological parameters such as contact rates that change in accordance with regional and state policies.

In addition to this,  some agent-based simulation features were added in the model which  better captures the characteristics of the pandemic. Using this model,  the following outputs for each U.S. state are created: i) the daily total number of COVID patients, ii) the total daily OC units to be acquired for hospitals and homes to support these patients, iii) current state-wide OC hospital inventory, iv) statewide number of OCs in use at hospitals, and v) the number of OC unit scrapped as a function of usage.  All of these signals are fed into the supply chain simulation model discussed in the next section (See Section \ref{sec:DES}).

On the starting date of the simulation, an initial number of COVID cases in each state is used based on the historical CDC data \shortcite{UnitedSt53}. It is assumed that a certain percentage of the total population is hospitalized after being infected by COVID and a certain percentage of this hospitalized population will also need OC support \shortcite{stasi2020treatment}.
In the SIR model,  five ``stocks'' that a person may be in at each time stage in a given U.S. state are defined.

\begin{enumerate}
    \item Susceptible: The group of people who are at risk of infection, a function of the total population, average infectivity rate, and a dynamic contact rate. 
    \item Infectious: Proportion of the population currently infected by COVID.
    \item Hospitalized: A system dynamic flow from Infectious, the patients hospitalized may also transition into a) recovered stage based on hospital recovery rate, or b) pass away. The model determines the total OC needs based on this stock.
    \item Deceased: The system dynamics flow from Infectious, the group of people who pass away due to either infection or after being hospitalized is represented at this stock which is the only stock without an outgoing flow.
    \item Recovered: The group of people who are recovered either come from the infectious state or COVID-based hospitalization. Importantly, recovered patients face a immunity loss after a fixed period and return to the susceptible state, capturing the possibility of reinfection.
\end{enumerate}

The user provides a list of states, total population \shortcite{StatePop99} and hospital capacity \shortcite{TotalHos43} for each state  for the baseline model. Although the average infectivity rate remains the same, the contact rates are dynamic and user-tunable to capture mitigation policies and seasonal infection rate changes. Note that the model is data driven and could accept any type of regions with their associated populations and epidemiological parameters to simulate. Table \ref{Table:1} reports the model details and input information for the baseline scenario that is defined as the original set of experiments without parameter tuning. 

\begin{table}[htbp]
\centering
\caption{SIR Model Parameter Details in the Baseline Scenario}
\setlength\tabcolsep{4pt} 
\resizebox{\columnwidth}{!}{
\begin{tabular}{lll} 
 \hline
 Parameter / Source & Value  & Details \\ 
 \hline
 Illness Duration \shortcite{tenforde2020symptom} & 15 days  & num of days before being recovered\\
 Simulation Time  & 11-20/3-21 & start and end date of the SIR model \\
 COVID Hosp. Rate \shortcite{menachemi2021many}& 0.01 &	pct of infected people hospitalized per day in each state \\
OC Inventory Rate &	0.10	& prop of OC usage held in stock at a hospital \\
OC Scrap Rate	& 0.01	& prop of OC scrapped per day at a hospital \\
COVID Hosp. Stay \shortcite{zeleke2022length} &  $X  \sim U[8,15]$ & num of days spent at a hospital due to COVID \\
% Non-COVID Hospitalization Rate & 0.1
%  & pct of other diseases per day in each state\\
% Non-COVID Hospitalization OC Rate & 0.1
%  & pct of non-COVID patients requiring OC support\\
 % Non-COVID Hospital Length of Stay &  $Y  \sim \exp(0.1)$ & num of days spent at a hospital due to non-COVID \\
Pct Population in Workforce
 &  0.5 & prop of population in the workforce\\
Immunity Duration \shortcite{Antibodi63}
 &  30 days & min num of days before reinfection\\
 % Invacare Market Share &	0.17 &	prop. of the U.S. market for Invacare \\
 OC Units Per Hospital Bed& 0.1 &  prop. of existing OCs in each state based on hospital beds\\
 Pre-COVID OC Demand Hospital  & 171 & num of OC needed before COVID at hospitals per day \\
  Pre-COVID OC Demand Home   & 545 & num of OC needed before COVID at homes per day \\
  OC Hospital COVID Usage  & 0.065 & prop. of COVID hospitalizations requiring an OC %, tuned using Maia Research Report and CDC data
  \\
OC Discharge COVID Usage  & 0.01 & prop. of discharges  requiring an OC after COVID%, tuned using Maia Research Report and CDC data
\\
OC Overflow Discharge COVID Usage  & 0.02 & prop. of overflow discharges requiring an OC
\\
\hline
\end{tabular}}
\label{Table:1} 
\end{table}

Both ``OC Inventory Rate'' and ``OC Scrap Rate'' will trigger the supply chain model to order OC unit in a hospital region. For example, if the number of OC units required is greater than (1- ``OC Inventory Rate'')$\%$, then the model will order ``OC Inventory Rate'' of the total demand to replenish the OC inventory. Since not having granular hospital inventory policies, this is a method for creating a consistent inventory policy across states.  In addition,  a couple of agent based simulation features is used within the system dynamics model. Each unit of flow is treated to the ``Hospitalized'' stock as an agent and sample from the user defined distribution individually what will be the length of stay, which changes dynamically when the Hospitalized population reaches the statewide capacity. Then keeping a memory structure  the flow is set to the ``Recovered'' stock accordingly. In a similar way, the direct flow from the ``Susceptible'' stock to the ''Recovered'' stock is being dictated with an agent-based approach. This approach makes the SIR model more realistic and allows for small spikes that sometimes diverge from the typical ''use the average'' system dynamics approach. 

\begin{figure}[htbp]
    \begin{minipage}{0.48\textwidth}
   \includegraphics[page=5,width=1\linewidth]{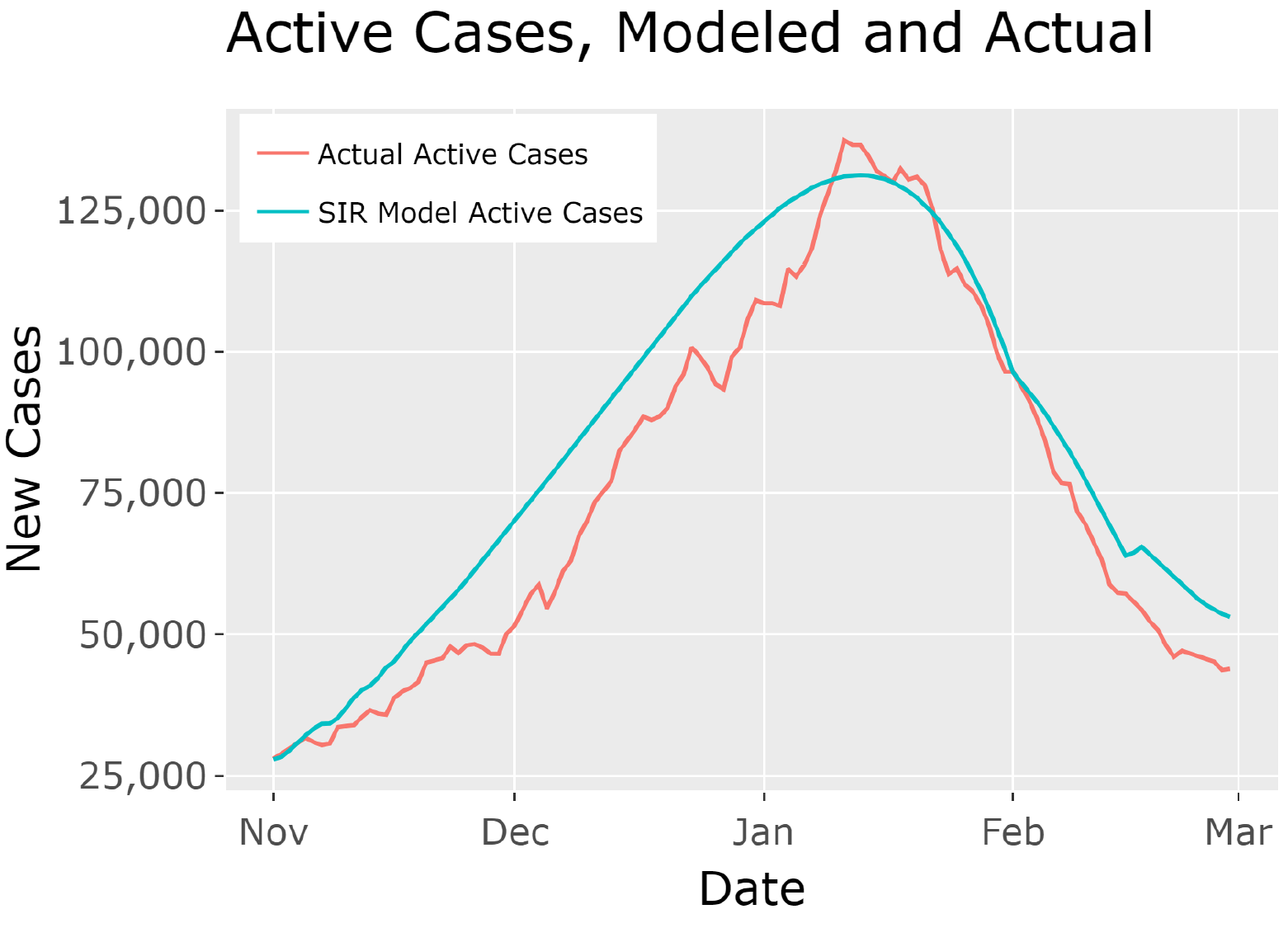}  
\caption{Comparisons of the active cases between the SIR model and actual in Georgia (GA)}
  \label{Fig6}
    \end{minipage}%
      \hfill
    \begin{minipage}{0.48\textwidth}
   \includegraphics[page=6, width=1\linewidth]{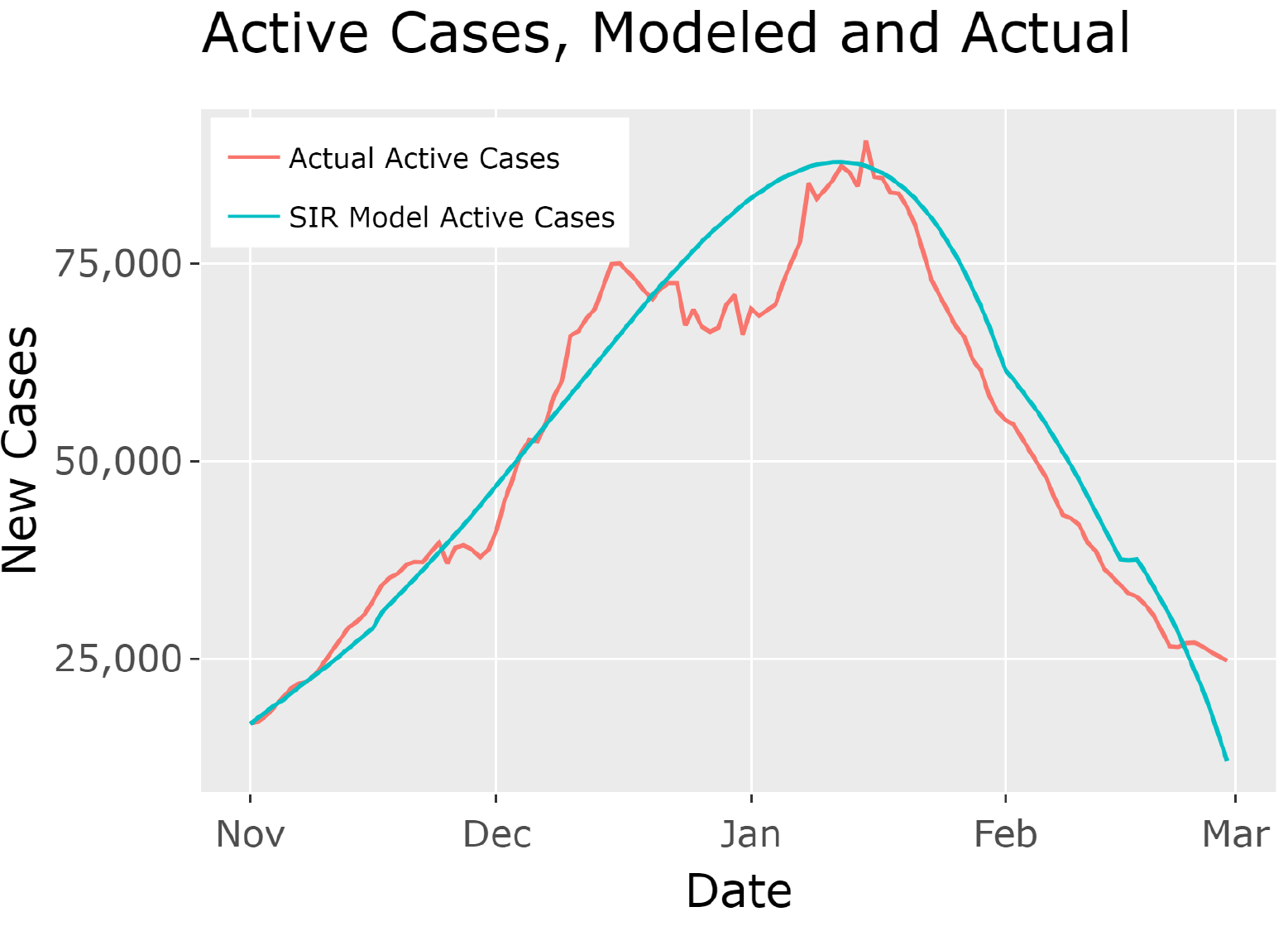}  
\caption{Comparisons of the active cases between the SIR model and actual in Massachusetts (MA)}
  \label{Fig7}
    \end{minipage}
\end{figure}

Another important parameter is ``Pct Population in Workforce'' where the Infectious stock is used to compute the number of people in the workforce infected by COVID. This result is used to incorporate degraded performance at manufacturers and suppliers due to workforce illness.  Further, both  ``Pre-COVID OC Demand Hospital'' and ``Pre-COVID OC Demand Home'' are used to capture baseline non-COVID related OC demand. These parameters are tuned based on national data from the ``Global Stationary Oxygen Concentrators Industry Market Research Report'' published by Maia Research \cite{maia}. Lastly,  daily COVID cases predicted by the SIR model (i.e., Infectious)  and actual COVID cases reported by CDC are compared across all modeled regions. Figures \ref{Fig6} and \ref{Fig7} compare the modeled COVID case count with actual case count for the example states of GA and MA respectively, demonstrating the accuracy of the modelling approach proposed.

\section{SUPPLY CHAIN DISCRETE EVENT SIMULATION} \label{sec:DES}

The OC supply chain model comprises of suppliers, main-assembly and sub-assembly facilities, OC products, assembly personnel, equipment, distributors, and customers. The multiscale simulation framework introduced by \shortcite{wang2019multiscale} based on the combination of discrete-event and agent-based simulation techniques is utilized  to model each entity's micro-scale activities within each manufacturing facility and macro-scale interactions at the supply chain network level. This framework enables the modeling of the individual components and the entire system, providing a comprehensive view of the OC supply chain and facilitating the identification of opportunities for optimization and improvement.

\subsection{Macroscale Model}

\begin{figure}[!htbp]
\centering
   \includegraphics[page=1,width=0.75\linewidth]{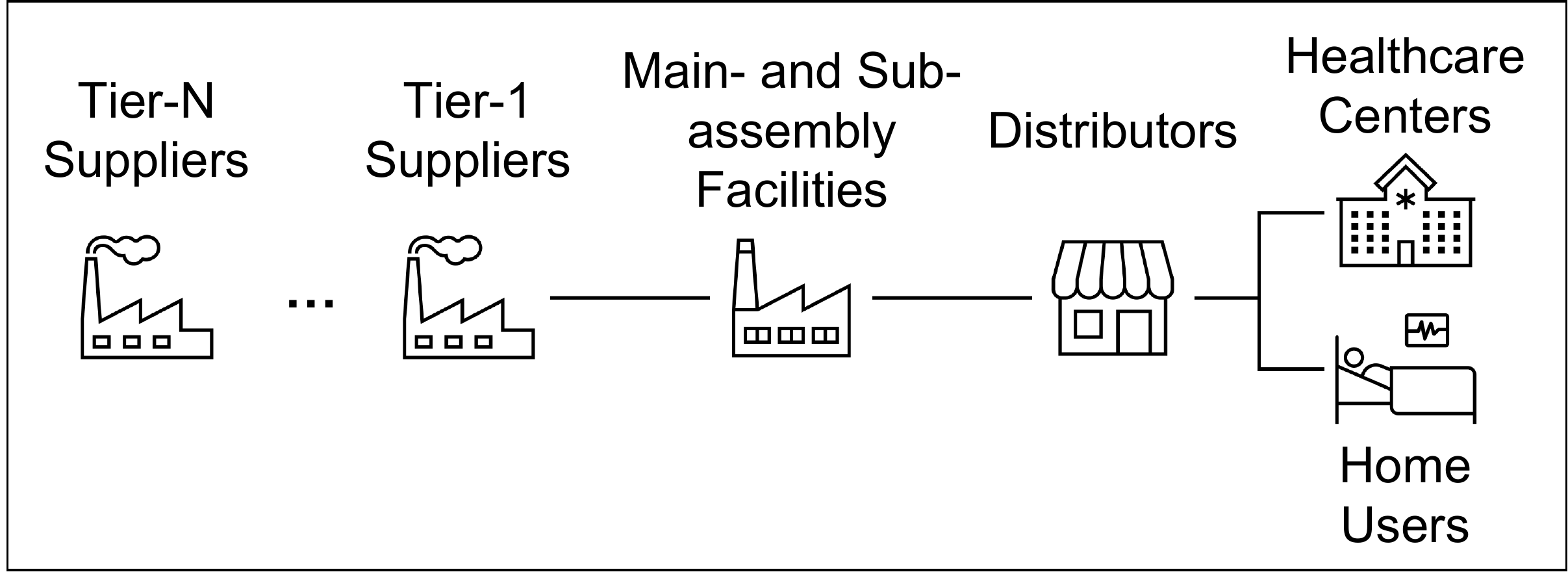}
\caption{Macroscale view of the OC supply chain in this study.}
  \label{Fig3}

\end{figure}

Figure~\ref{Fig3} shows the relationships between entities within the OC supply chain. The macroscale model simulates the OC supply chain network and associated activities, which include the allocation of distributors, assembly facilities, suppliers, and transportation activities. %Users can configure the number and location of the entities and their relationships by importing data through SQL databases or Excel spreadsheets. 
In addition, the macro-scale simulation model has an integrated geographic information system (GIS) map that allows users to watch the animation of agent activities and interact with the simulation by clicking on any entities to access its parameters or microscale model presentation. %The GIS map integrated with the OC supply chain simulation model is illustrated in Figure~\ref{Fig4}. 
The OC supply chain considered in this study contains 52 OC distributors, one assembly facility, and 278 suppliers across the globe. To gather this data, RADAR utilized a Deloitte-led supplier illumination to provide a comprehensive understanding of the location, industry, and relationships of these 278 suppliers within the supply chain network.
% \begin{figure}[htbp]
% \centering
%    \includegraphics[page=2, width=0.75\linewidth]{gt.pdf}  
% \caption{Example view of the GIS map of the OC supply chain model}
%   \label{Fig4}
% \end{figure}

% \begin{figure}[htbp]
%     \begin{minipage}{0.48\textwidth}
%    \includegraphics[page=1,width=1\linewidth]{gt.pdf}  
% \caption{Macroscale view of the OC supply chain in this study}
%   \label{Fig3}
%     \end{minipage}%
%       \hfill
%     \begin{minipage}{0.48\textwidth}
%    \includegraphics[page=2, width=1\linewidth]{gt.pdf}  
% \caption{Example view of the GIS map of the OC supply chain model}
%   \label{Fig4}
%     \end{minipage}
% \end{figure}
\subsection{Microscale Model}
The microscale model simulates activities within distributors, main-assembly facilities, and sub-assembly facilities. Within each entity of the considered OC supply chain, there is a set of static or stochastic attributes that reflect the entity’s present condition, described below. %A representative process flow of the OC supply chain at the microscale level is illustrated in Figure~\ref{Fig5}.

% \begin{figure}[htbp]
% \includegraphics[page=1,width=1\linewidth]{Figure3.pdf}  
% \caption{Microsacle  view of the OC supply chain in this study}
%   \label{Fig5}
% \end{figure}

The OC distributor order process consists of these steps: First, the distributor receives an order for a specified number of OCs from a healthcare facility or a customer. The distributor checks inventory to verify that the requested OC models are available in the desired quantity. Additionally, for individual customers, the distributor verifies prescription details with the relevant healthcare providers. Once the details have been confirmed, billing and payment processes are initiated, and the OCs are packaged and prepared for shipping to the designated location. If the distributor's inventory is insufficient, a new order is added to the customer queue to wait for inventory replenishment. To maintain inventory level, a standard continuous review $(Q,R)$ inventory policy \shortcite{eksioglu2008highway} is implemented to monitor and make replenishment orders of quantity $Q$ to the assembly facility when the inventory level is below the reorder point $R$, assuming the expected cycle period is $T$, and the weekly demand has a normal distribution with mean $\mu_D$ and standard deviation $\sigma_D$. The lead time for inventory replenishment follows a normal distribution with mean $\mu_{(L_T)}$ and standard deviation $\sigma_{(L_T)}$. The equations for calculating $Q$ and $R$ are expressed in (\ref{eq}).
\begin{align} \label{eq}
\begin{split}
Q &= \mu_D  T \\
R &= \mu_D \mu_{L_T} +Z_{\alpha}  
\end{split}
\end{align}

% $$Q = \mu_D  T $$ $$R = \mu_D \mu_{L_T} +Z_{\alpha}  \sqrt{\mu_{L_T} \sigma^{2}_D  + \sigma^{2}_D  \sigma^{2}_{L_T}}$$

 When an order from the distributor is received at an assembly facility, the assembly OC inventory is checked, and if there is sufficient inventory, the OCs are prepared for shipping to the distributor. However, if the stock is insufficient, the order enters a queue to wait for newly assembled OCs. Similar to the distributor, a continuous review $(Q,R)$ inventory policy is implemented to monitor the OC inventory at the assembly facility. Assembly jobs are created within the facility to produce OCs when inventory levels fall below the reorder point. The main assembly process flow consists of four primary steps, each assembling sub-assembly parts into the main OC body. The microscale model also includes sub-assembly facilities for producing and supplying sub-assembly parts. An inventory policy is also implemented at the sub-assembly facilities to control material inventory levels.

\section{COMPUTATIONAL EXPERIMENTS} \label{sec:Experiments}

% In this section, we discuss the computational experiments. We note that the SIR model experiments are conducted using the Java API and AnyLogic library 9.5.1 on a computer
% having an 11th Gen Intel Core i7-11850H processor and 32 GB of RAM. As for the supply chain model, we perform our experiments using the Java API and AnyLogic library 8.7.7 on a desktop PC with 2.40 GHz, Intel(R) Core (TM) i7-9700K CPU, 16.0 GB RAM.

The computational experiments were conducted  using the Java API \shortcite{camur2021optimizing} and AnyLogic library.%. The SIR model was run on a laptop with an 11th Gen Intel Core i7-11850H processor and 32 GB of RAM. The supply chain model was run on a desktop PC with an Intel(R) Core (TM) i7-9700K CPU, 16.0 GB RAM, and 2.40 GHz. 

\subsection{SIR Model Experiments} 

In this section,  certain parameters presented in Section \ref{sec:SIR}  are tuned to perform what-if analyses. These results will then be used to further analyze the supply chain model proposed in the following section. In both sections, ten states are focused on: AK, AZ, CA, GA, IL, MA, SD, VT, WI, and WY to scale the analysis. These states were chosen because they display either a large case count due to population, or a high sensitivity to model parameter changes. The time-variant contact rate information is increased by $0.1\%$ aiming to observe the impact of a more aggressive pandemic behavior on the population and total OC demand. SIR models can be quite sensitive to small changes in contact rates for high infectivity viruses, resulting in exponential increases in infectious rates depending on the current state of the susceptible, infectious and recovered populations.

\begin{figure}[!htbp]
\centering
\includegraphics[scale=0.25]{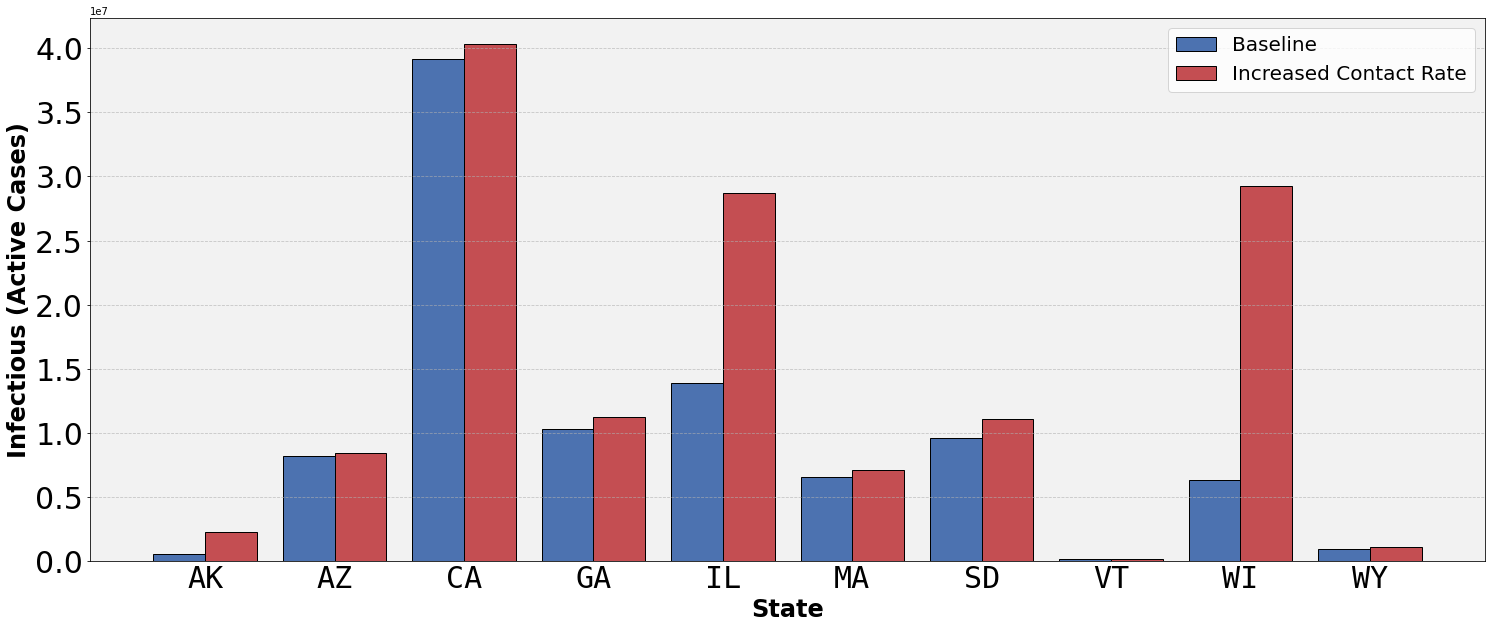}
\caption{Comparison of infectious values per state between the baseline  and  increased contact rate scenarios.}
\label{Fig5}
\end{figure}

The comparison of the total number of people infected in each state between the baseline and increased contact rate scenarios is presented in Figure ~\ref{Fig5}. Several states (e.g., AK, IL and WI) indicate a high sensitivity to increases in contact rate, particularly after infection peak. This sensitivity is a function of the previous infections in that state and the status of the recovered population. More sensitive states have a lower proportion of their population in the recovered category of the SIR model. Therefore, states will have varying sensitivities to increases in reproduction numbers based on the current infected population, recovered population, etc. In essence, for those states, an endemic situation can turn into a pandemic wave with the increased contact rates. Figure ~\ref{Fig5}  illustrates that each state shows a unique behavior during the pandemic based on varying contact rates, mitigation policies, underlying weather, neighboring state behavior, and population health, thus, analyzing the results as a whole may not provide useful insights. 

\begin{figure}[!htbp]
     \centering
   \includegraphics[page=1,width=1\linewidth]{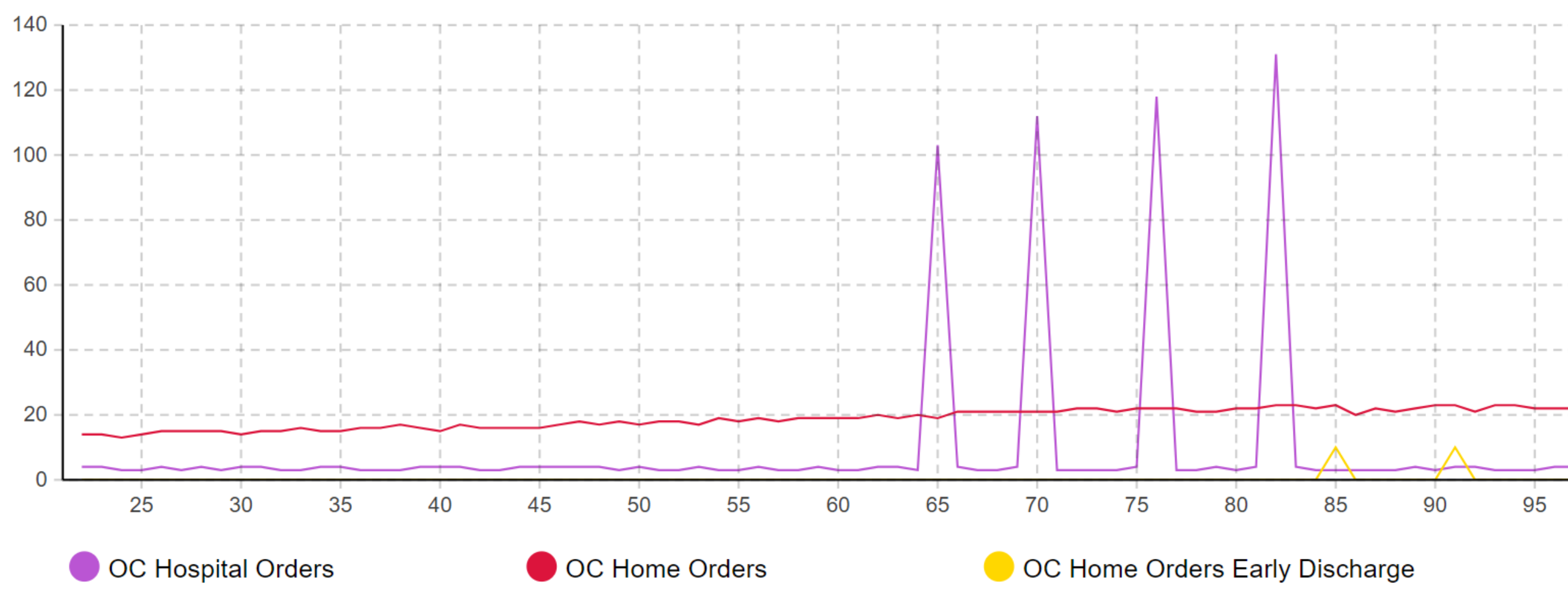}  
\caption{OC orders in AZ in the baseline experiment over time in days (x-axis).}
  \label{Fig8}
\end{figure}

\begin{figure}[!htbp]
     \centering
   \includegraphics[page=2, width=1\linewidth]
   {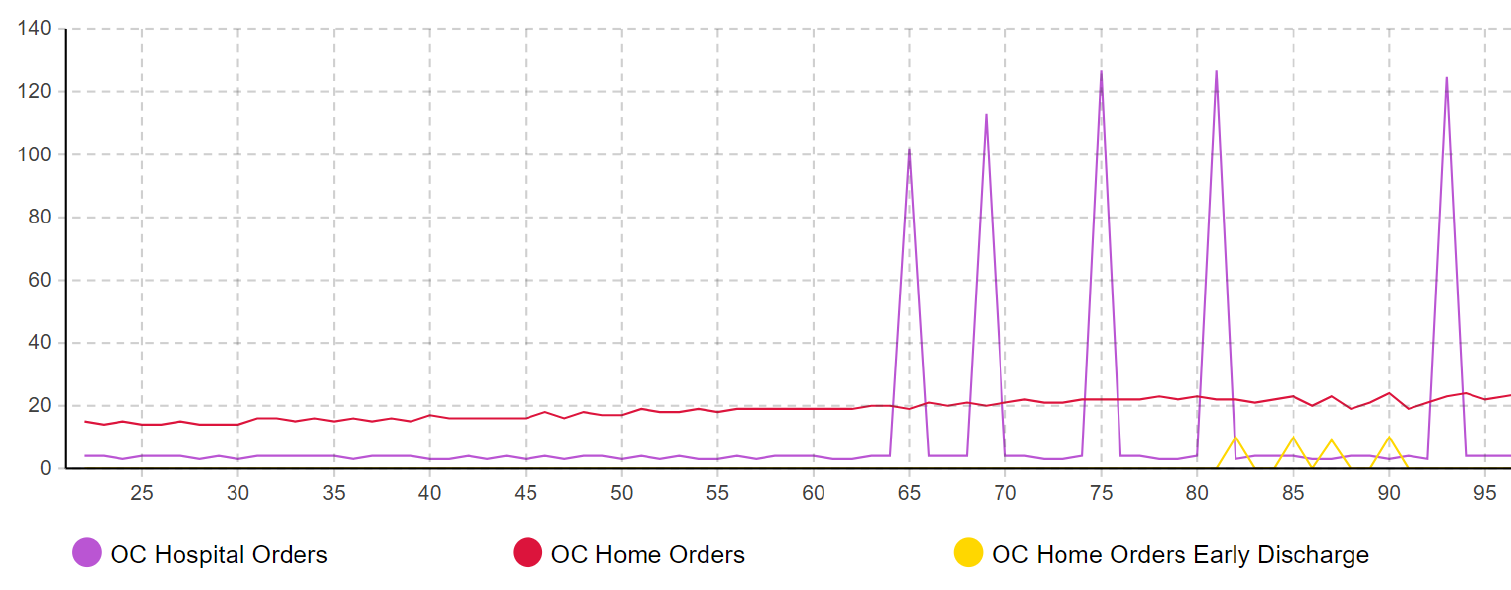}  
\caption{OC orders in AZ in the increased contact rate experiment over time in days (x-axis).}
  \label{Fig9}
\end{figure}

To gain a deeper understanding of impact in a certain state (i.e., AZ),  OC orders/demands are analyzed. In the baseline scenario, it is found that the number of hospitalized patients exceeded total hospital capacity (i.e., 14k) twice in AZ, which triggered two early OC home orders on days 85 and 92 (see Figure \ref{Fig8}). The analysis reveals that the contact rate plays a significant role in driving the infection and hospitalization rates specifically during the peak of the pandemic, resulting in double the number of early OC home orders once the contact rate is modified, even by a small scalar. It is observed that the total number of available OCs goes below the minimum inventory limit $5$ times, leading the model to place an average of 117 OC orders $5$ times between the second and third months, as shown in Figure ~\ref{Fig9}.

\begin{figure}[H]
    \begin{minipage}{0.48\textwidth}
    \centering
   \includegraphics[page=2,width=0.8\linewidth]{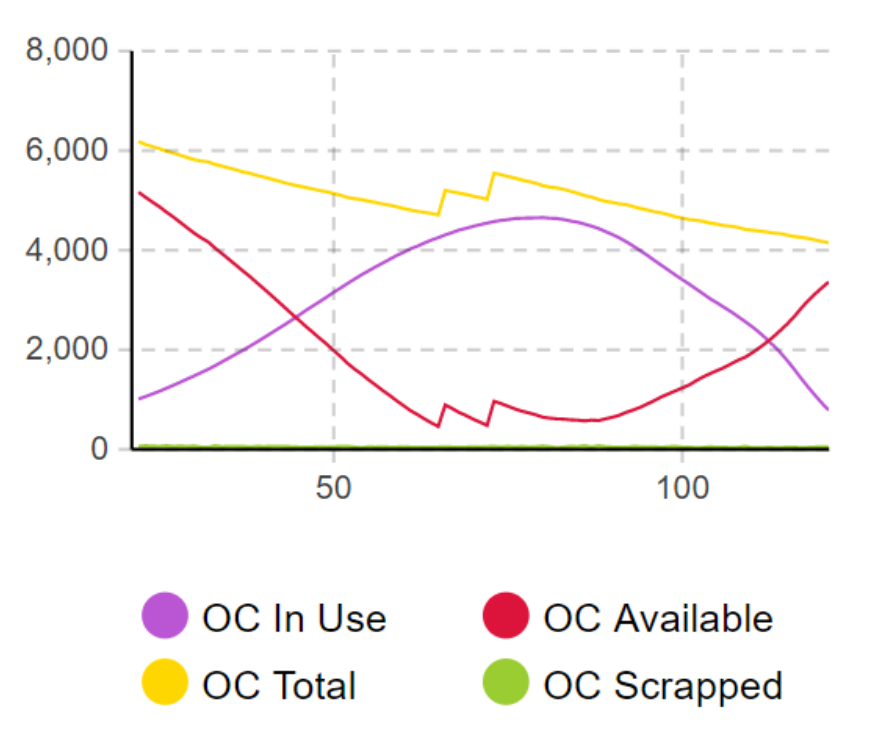}  
\caption{OC availability in CA in the baseline experiment over time in days (x-axis).}
  \label{Fig10}
    \end{minipage}%
      \hfill
    \begin{minipage}{0.48\textwidth}
     \centering
   \includegraphics[page=5, width=0.8\linewidth]{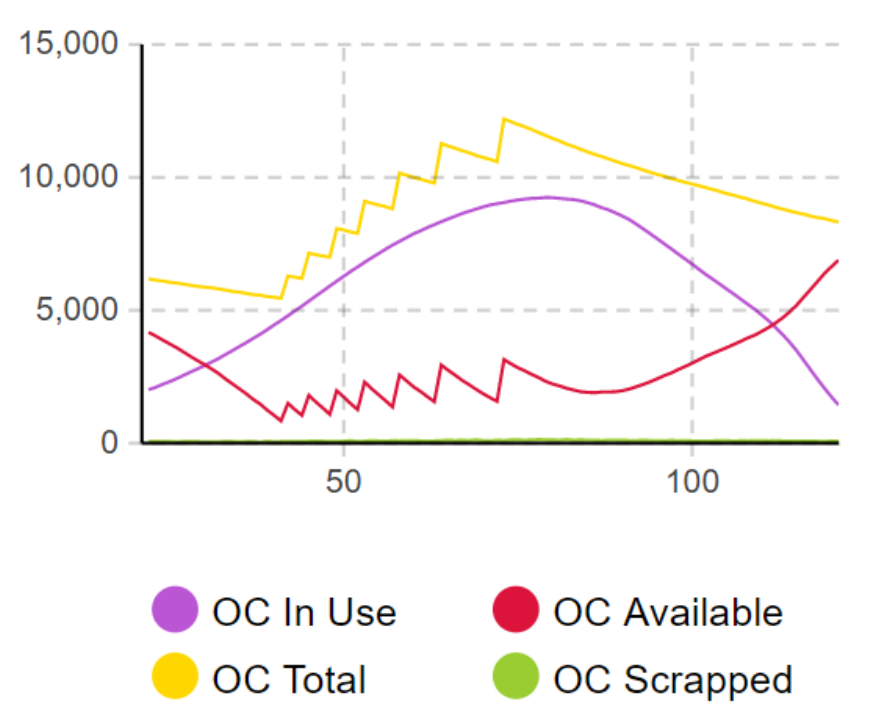}  
\caption{OC availability in CA in the increased OC usage rate experiment over time in days (x-axis).}
  \label{Fig11}
    \end{minipage}
\end{figure}

In another set of experiments,  a more impactful pandemic situation where COVID-related OC usage rates are doubled is tested (see Table \ref{Table:1}). Also,  the OC Inventory Rate is increased to $0.15$ to model a more cautious OC blanket inventory policy at hospitals.  For this experiment, the attention is turned to the state of California (CA). Since OC usage is higher for COVID patients, OC demand shows an exponential increase trend (Figure ~\ref{Fig11}), whereas,  a decrease in the baseline experiment is observed as illustrated in Figure ~\ref{Fig10} indicating that CA is able to handle the increased demand during the early stages. However, increased OC usage causes a cascading effect with OC hospital orders experiencing an exponential increase (an increase of $100\%$). In the baseline scenario, there are only $2$ large OC hospital orders with an average of 550 OCs, while this scenario requires $7$ inventory replenishments at the hospital level with 1,000 OCs on average.

\subsection{Supply Chain Model Experiments}
\begin{figure}[htbp]
\centering
\includegraphics[scale=0.55]{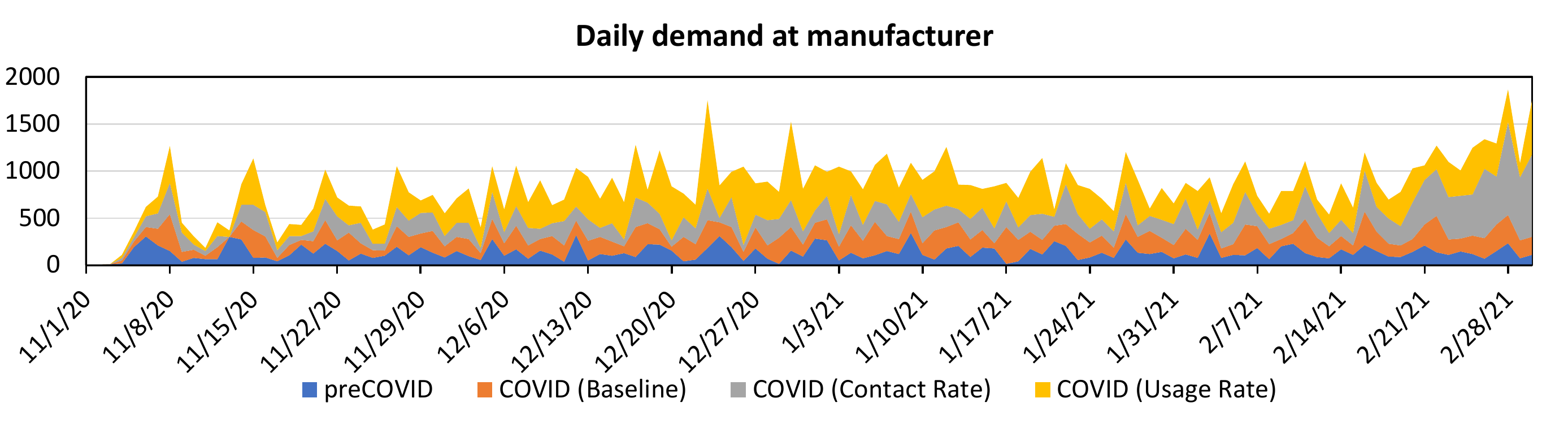}
\caption{Impact of COVID-19 pandemic on the daily demand at OC manufacturer.}
\label{Fig12}
\end{figure}

In this section, the impact of the OC demand surge on supply chain performance during the COVID-19 pandemic is investigated by evaluating the effectiveness of mitigation strategies, particularly dynamic inventory policies and the incorporation of air freight as a mode change from surface freight transportation. A pre-COVID-19 demand scenario as well as three COVID-19 demand scenarios created in the previous section: (i) baseline, (ii) increased contact rate, and (iii) increased usage rate are considered. In the previous section, it is indicated that each COVID-19 scenario leads to a surge in demand across numerous states. Consequently, the heightened demand for regional distributors translates to an elevated daily demand for the OC manufacturer, as depicted in Figure ~\ref{Fig12}. When compared to the pre-COVID-19 period, the average daily demand increase for scenarios (i), (ii), and (iii) amount to 33\%, 47\%, and 78\%, respectively.
%Figure ~\ref{Fig12}(a) presents the total demand for each regional distributor under various demand scenarios, considering the possible shifting hotspots from one state to another over time. The data demonstrate that each COVID-19 scenario leads to a surge in demand across numerous states. Consequently, the heightened demand for regional distributors translates to an elevated daily demand for the OC manufacturer, as depicted in Figure ~\ref{Fig12}(b). When compared to the pre-COVID-19 period, the average daily demand increase for scenarios (i), (ii), and (iii) amount to 33\%, 47\%, and 78\%, respectively.

The initial analysis focuses on the impact of the pandemic on lead times for customer orders and inventory replenishment orders by distributors. It is assumed that manufacturers and distributors are following their pre-pandemic inventory management policies, which aim to achieve a service level of 95\%. Additionally, it is assumed the manufacturer has sufficient material supply and transportation capabilities and uses ground transportation to deliver OC to the distributors. Finally, it is assumed that the manufacturer has limited labor and workspace available, which were utilized at a 50\% rate before the pandemic.

\begin{figure}[htbp]
\centering
\includegraphics[scale=0.46]{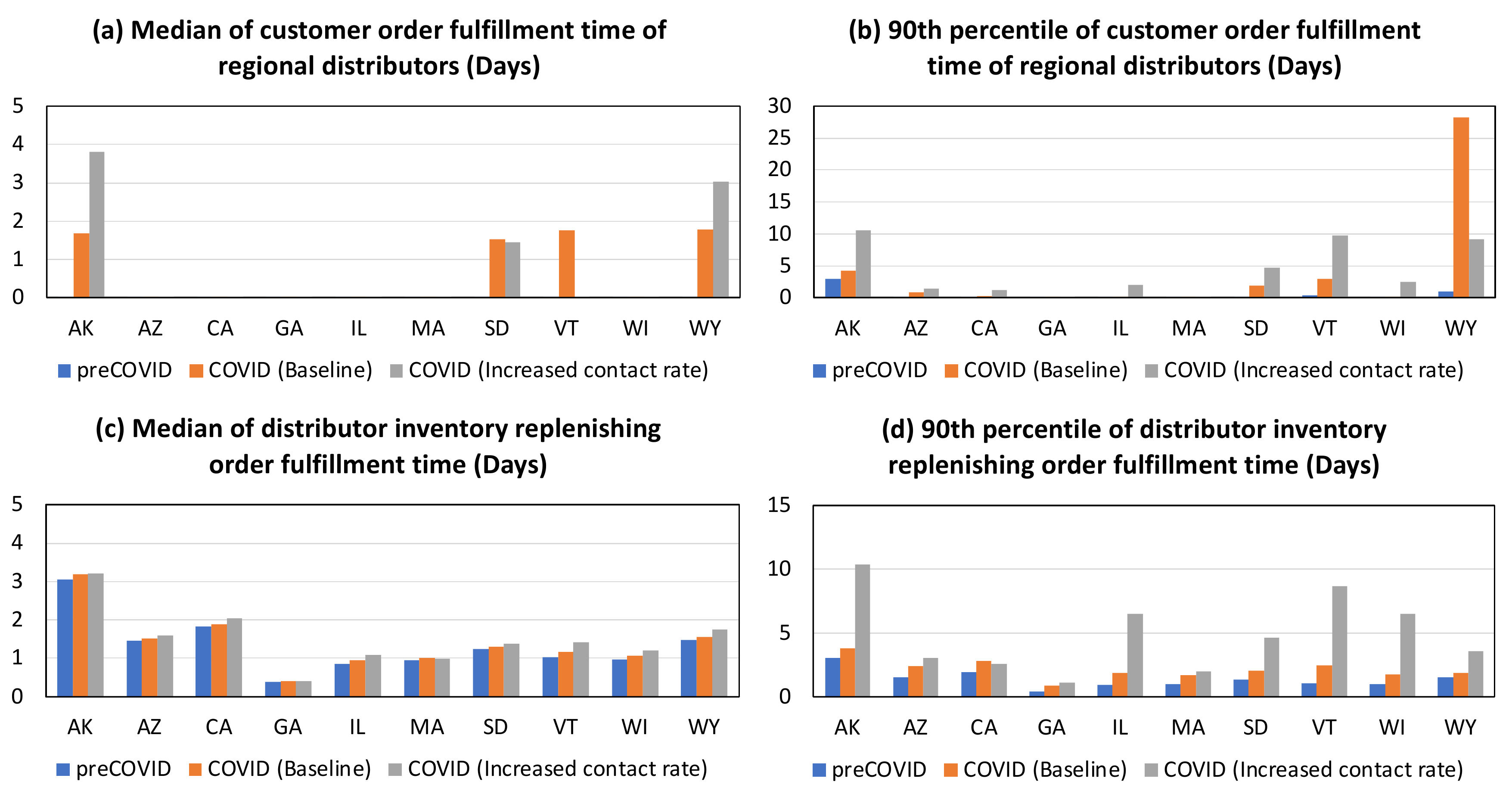}
\caption{Fulfillment time statistics of customer orders.}
\label{Fig13}
\end{figure}

Figure ~\ref{Fig13} presents the lead time statistics for ten selected regional distributors. %hese states are selected due to their notable increase in lead times. 
Note that the demand scenario (iii) is not displayed in Figure ~\ref{Fig13} due to its exceedingly long lead times for all states, e.g., the median and 90$^{th}$ percentile customer fulfillment time are 8 and 38 days. Figure~\ref{Fig13} (a) and (b) demonstrate a significant increase in customer fulfillment times for AK, SD, VT, and WY. This finding is noteworthy as it indicates a potential issue with the efficiency of inventory replenishment in these specific regional distributors, to be mitigated by other means. %Notably, the customer fulfillment times at AK, SD, VT, and WY increase significantly, as shown in Figure ~\ref{Fig13}(a) and (b). 

\begin{figure}[htbp]
\centering
\includegraphics[scale=0.5]{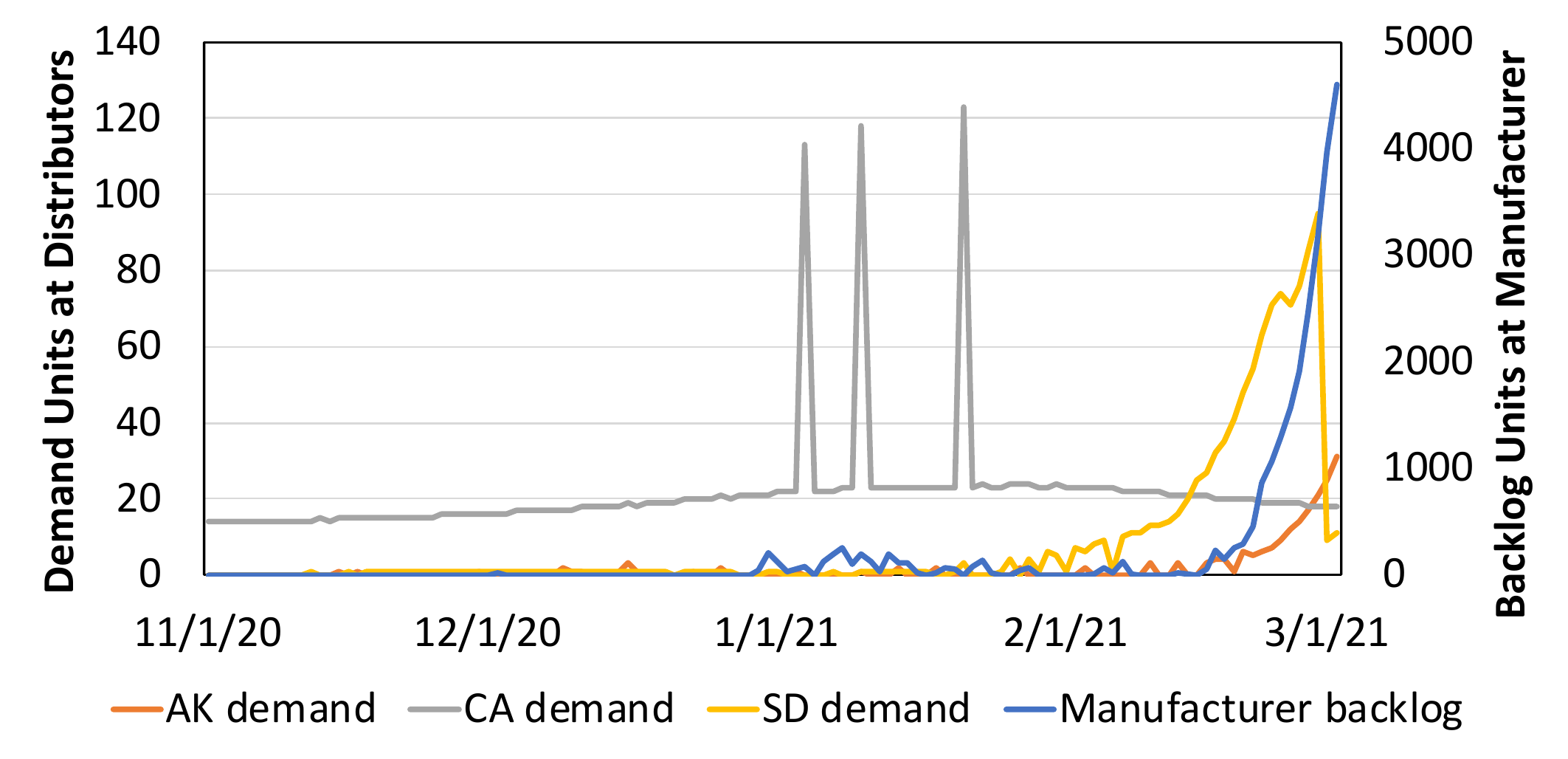}
\caption{Daily demand at distributors vs. manufacturer backlog (increased contact rate scenario).}
\label{Fig14}
\end{figure}

To better understand the lead time increases in the selected states, the time series of daily demand at AK, CA, and SD distributors, along with the backlog level at the manufacturer are analyzed, as illustrated in Figure~\ref{Fig14}. The analysis reveals that the demand spikes at SD and AK distributors coincide with periods of high backlog levels at the manufacturer, which is likely the primary reason for increased customer fulfillment times. Figure~\ref{Fig14} also indicates that the CA distributor experienced three demand spikes before AK and SD demand spikes, and before the manufacturer backlog occurred. Because of this timing difference, the CA distributor did not experience a large rise in customer fulfillment time (see Figure ~\ref{Fig13}), but these large orders depleted stocks at the manufacturer, leading to larger backlogs. Because small states such as VT and WY have fewer orders, their lead time statistics and inventory policies are more susceptible to extreme values. For example, the order-up-to-inventory levels are $2$ for AK, VT, and WY and $3$ for SD. Introducing an additional stock in small regions for emergencies could mitigate risks at low total cost.

Further, two mitigation strategies and their potential advantages are examined. The first strategy involves adjusting inventory policies dynamically based on the previous week's demand, while the second strategy incorporates dynamic inventory policies and permits manufacturers to transport OCs to distributors via air freight when the distance between them exceeds 500 miles.
To summarize the results of the evaluation, Table~\ref{Table:2} presents the fulfillment time of these mitigation strategies under different demand scenarios. The findings indicate that adjusting inventory policies based on demand data can significantly reduce lead time. Furthermore, air freight can reduce fulfillment time for distributor replenishment orders in the baseline and increased contact rate scenarios. An increase is observed in the 90$^{th}$ percentile manufacturer's fulfillment time under increased usage rate for distributor replenishment orders. The preliminary findings suggest this is due to the enlarged size of distributors' replenishment orders under dynamic inventory policies. While the implementation of dynamic inventory policies or the utilization of air freight has the potential to reduce lead time, it is crucial to consider additional costs or bottlenecks arising from implementing these strategies. Further studies are necessary to gain deeper insights into these potential effects.

\begin{table}[htbp]
\centering
\caption{Fulfillment time summary of customer and distributor replenishment orders.}
\setlength\tabcolsep{4pt} 
% \resizebox{\textwidth}{!}{
\begin{tabular}{lcccc}
\hline
\multicolumn{5}{c}{90$^{th}$ percentile fulfillment time for national customer order (Days)} \\
\hline
Demand scenario & pre COVID-19  & Baseline & Increased contact rate & Increased usage rate \\ 
\hline
%\multicolumn{1}{p{4cm}}{\raggedright Static inventory + Ground transport} & 0.012 & 0.021 & 0.021 & 6.958\\
Static + Ground & 0.02 & 0.04 & 0.4 & 38.6\\
Dynamic + Ground & 0.02 & 0.04 & 0.04 & 37.4\\
Dynamic + Air & 0.02 & 0.04 & 0.04 & 37.3\\
\hline
\multicolumn{5}{c}{90$^{th}$ percentile fulfillment time for national distributor replenishment orders (Days)} \\
\hline
Demand scenario & pre COVID-19  & Baseline & Increased contact rate & Increased usage rate \\ 
\hline
Static + Ground & 1.8 & 2.1 & 3.3 & 41.2\\
Dynamic + Ground & 1.9 & 2 & 3.1 & 44.7\\
Dynamic + Air & 0.9 & 1.2 & 1.7 & 43.7\\
\hline
\end{tabular}
\label{Table:2} 
\end{table}

\section{CONCLUSION} \label{sec:Conclusion}
In this study,  a novel hybrid simulation framework combining a SIR model and a discrete event simulation is proposed to understand the demand pattern of oxygen concentrators (OCs) and identify how to meet nonstationary demand during a disruptive event, like a pandemic.  A pre-COVID-19 demand scenario and  three COVID-19 demand scenarios using the SIR model: (i) baseline, (ii) increased contact rate, and (iii) increased OC usage rate are investigated. The supply chain model proposed then examines the impact of these different demand signals, determining the sensitivity to changes in inventory and transportation policy to reduce lead times during the supply and demand shock.

The integrated simulation framework is shown to be robust and valid using the real-world data provided by government and research organizations (i.e., CDC, Maia Research). The model incorporates regional and demographic effects of COVID to enable a supply chain manager to pinpoint specific regions or states which will suffer comparatively more from a supply chain shock, and enables the rapid testing of mitigation policies (e.g., dynamic inventory, air freight) to determine how to minimize the effects of these shocks.  These policies can then be deployed at a granular, regional level, enabling a more robust shock response while minimizing cost and resource usage. The team plans to incorporate additional scenarios into this framework, including the targeted incorporation of robotics and automation technologies in the supply chain to further reduce lead times and improve response to these shocks.  This work was supported by the U.S. Department of Commerce under grant number 70NANB22H012.

% Reducing font size (to 9pt) for References & Author Biagraphies
\footnotesize

% Please don't exchange the bibliographystyle style
\bibliographystyle{wsc}

% AUTHOR: Include your bib file here
% \bibliography{demobib}

\section*{AUTHOR BIOGRAPHIES}

\noindent {\bf MUSTAFA C. CAMUR} works as a research engineer in the Optimization $\&$ Risk Analysis  Team  at GE Research (GER).  He received his Ph.D. degree at Clemson University in Industrial Engineering. His research interests include network optimization, decomposition algorithms, and applied machine learning. His e-mail address is \email{can.camur@ge.com}.\\

\noindent {\bf CHIN-YUAN TSENG} is a Ph.D. candidate in Industrial Engineering  at Georgia Institute of Technology, focusing on simulation, reinforcement learning and dynamic optimal control theories for production systems and supply chain integration. His e-mail address is \email{ctseng40@gatech.edu}.\\

\noindent {\bf ARISTOTELIS E. THANOS} works as a Senior Lead Engineer in the Optimization $\&$ Risk Analysis Team at GER.  He received his Ph.D. degree in the Department of Industrial Engineering at the University of Miami. His research is on simulation and optimization of large-scale systems with a focus on energy and its applications. His e-mail address is \email{a.thanos@ge.com}.\\ 

\noindent {\bf CHELSEA C. WHITE III} holds the Schneider National Chair of Transportation and Logistics at Georgia Institute of Technology.  His research interests include analyzing the role of real-time information for stress testing supply chains to improve next-generation supply chain competitiveness and resilience. He can be reached at \email{w196@gatech.edu}\\

\noindent {\bf WALTER YUND} works as a Senior Scientist in the Optimization $\&$ Risk Analysis Team at GER.  He received his Ph.D. degree in the Department of Industrial and Systems Engineering at Rensselaer Polytechnic Institute. His research interest is in resilient supply chain management and data analytics. His e-mail address is \email{yund@ge.com}\\

\noindent {\bf ELEFTHERIOS IAKOVOU} is the Harvey Hubbell Professor of Industrial Distribution at Texas  A$\&$M University, and the associate director of Resilient and Sustainable Supply Chains for the Texas A$\&$M Energy Institute. He holds M.Sc. and Ph.D. degrees from Cornell University.  His e-mail address is \email{eiakovou@tamu.edu}\\

\end{document}